\documentclass[prb,twocolumn,showpacs]{revtex4-1}
\usepackage{graphicx}
\usepackage{multirow}
\usepackage{dcolumn}
\usepackage{bm}
\usepackage[normalem]{ulem}
\usepackage{amsmath}
\usepackage{amsfonts}
\usepackage{amssymb}
\usepackage{color}
\usepackage{colordvi}
\usepackage{dcolumn}
\usepackage{subfigure}

\newcommand{\br}{\mathbf{r}}

\newcommand{\bq}{\mathbf{q}}
\newcommand{\bp}{\mathbf{p}}
\newcommand{\bk}{\mathbf{k}}
\newcommand{\bpp}{\mathbf{p}^\prime}

\newcommand{\bB}{\mathbf{B}}

\newcommand{\bD}{\mathbf{D}}

\newcommand{\bn}{\mathbf{n}}

\newcommand{\ii}{\mathrm{i}}

\newcommand{\pd}{\partial}

\newcommand{\ga}{\gamma}
\newcommand{\de}{\delta}

\newcommand{\w}{\omega}
\newcommand{\si}{\sigma}
\newcommand{\ep}{\varepsilon}
\newcommand{\epp}{\varepsilon^{\prime}}
\newcommand{\la}{\lambda}
\newcommand{\bphi}{\boldsymbol{\phi}}
\newcommand{\bsi}{\boldsymbol{\sigma}}

\begin{document}

\title{Pairing gaps near ferromagnetic quantum critical points}
\author{M. Einenkel$^{1,2}$, H. Meier$^{3}$, C. P\'epin$^{4}$, and K. B. Efetov$^{1,2}$}
\affiliation{
$^{1}$Institut f\"ur Theoretische Physik III, Ruhr-Universit\"at Bochum, 44780 Bochum, Germany
\\
$^{2}$National University of Science and Technology \textquotedblleft
MISiS\textquotedblright, Moscow 119049, Russia
\\
$^{3}$Department of Physics, Yale University, New Haven, Connecticut 06520, USA
\\
$^{4}$IPhT, L`Orme des Merisiers, CEA-Saclay, 91191 Gif-sur-Yvette, France}
\date{\today }

\begin{abstract}
We address the quantum-critical behavior of two-dimensional itinerant
ferromagnetic systems described by spin-fermion models in which fermions
interact with close-to-critical bosonic modes. We consider Heisenberg
ferromagnets, Ising ferromagnets, and the Ising nematic transition. Mean-field theory close to the quantum critical point predicts a superconducting
gap with spin-triplet symmetry for the ferromagnetic systems and a singlet
gap for the nematic scenario. Studying fluctuations in this ordered phase
using a nonlinear sigma model, we find that these fluctuations are not
suppressed by any small parameter. As a result, we find that a
superconducting quasi-long-range order is still possible in the Ising-like
models but long-range order is destroyed in Heisenberg ferromagnets.
\end{abstract}

\maketitle

\section{Introduction}

Quantum criticality and quantum critical points (QCP) are among the most interesting subjects of contemporary condensed matter physics from both theoretical and experimental points of view alike.\cite{Sachdevbook,Wolfle}
Here, we focus on the effects of quantum-critical behavior of two-dimensional (2D) itinerant ferromagnets and, in particular, the cases of Heisenberg and Ising ferromagnets. Also, we address the transition towards Ising nematic order. 

The Stoner transition towards ferromagnetic order can be regarded as an archetype of quantum phase transitions. Generally, it belongs to a class of transitions where the order parameter fields carries an ordering wave vector $\bq = 0$. These and similar systems like spin liquids and fermions coupled to a $\mathrm{U}(1)$-gauge field\cite{Lee,Altshuler} are suitably described in terms of $2+1$-dimensional field theories. In these theories, the fluctuations of the order parameter are Landau-damped bosons with the well-known propagator
\begin{align}
\chi(\w,\bq) = \frac{1}{\gamma |\w/q|+ \bq^2 + a}\ ,
\end{align}
where $\w$ denotes frequency and $q=|\bq|$. At the QCP, the boson mass $a$ turns to zero, so that frequencies scale with momentum as~$\w\sim q^z$ with a dynamical exponent~$z=3$. Close to this point, the low-energy behavior of fermions that interact with the fluctuations of the order parameter is driven away from Fermi-liquid theory, resulting in non-Fermi-liquid~(nFL) physics with different scaling relations.\cite{Altshuler,Lee, MetSachI, Rech, Rech2}

This strong interaction between critical bosons and massless fermions is essential for the description of the quantum-critical phenomena and its full impact has been missed in the earliest theoretical studies on this subject,\cite{hertz,millis} which predicted that critical behavior is mean-field-like in both two and three dimensions. In these works, the fermions were fully integrated out from the partition function, which led to a critical $\phi^4$~theory for the order parameter fluctuations alone. In two dimensions, such an
approach breaks down\cite{Wolfle, Rech} as the coupling to the fermions leads to nonanalyticities or even singularities in the $\phi^4$ theory. This suggested that
magnetic transitions are properly studied only in terms of so-called spin-fermion models,\cite{abchu1,abchu2,Rech,Rech2} which explicitly keep the interaction between fermions and order parameter fluctuations.

Spin-fermion models themselves caused their own troubles related to the question of analytical control of the theory. Introducing an artificially large number~$N$
of fermion ``flavors'' in addition to the two electronic spin species,\cite{Altshuler, Rech} a controlled solution was sought in the limit of $N \rightarrow \infty$ with perturbative corrections small in $1/N$. However, as discovered shortly after,\cite{Lee,MetSachI,MetSachII} a conventional $1/N$~expansion breaks down for these models as certain classes of Feynman diagrams
that by first-sight inspection seem small in $1/N$ in fact are not. Their neglect in the previous work is thus not justified. More severely, it is
unclear if the theory can be analytically controlled at all.

In this situation, several of us tried a different approach\cite{emp} to the spin-fermion model for an \emph{anti}ferromagnetic QCP, which was inspired by certain similarities between the ``dangerous'' diagrams mentioned above and those diagrams giving rise to the diffusion modes in the theory of localization.\cite{Ebook} In this approach, the dangerous diagrams are effectively summed and captured in terms of an effective saddle-point theory whose fluctuations are properly described in terms of a nonlinear $\sigma$~model.
It is then only natural to ask whether such a scheme may be applied to the ferromagnetic case and similar systems.

In the ferromagnetic case, even the existence of the QCP is under debate.\cite{Rech,Rech2,Belitz1,Kirkpatrick,Belitz2,Karahasanovic,Conduit,Pedder,Kruger}
The authors of Ref.~\onlinecite{Rech} considered an itinerant ferromagnet in the non-Fermi-liquid regime and showed that non-analytic contributions to the
static particle-hole susceptibility~$\chi(0,\bq)$ appear for the Heisenberg ferromagnet and most likely destroy the nature of the second-order phase transition in this system.
Only at low temperatures they expected a first-order transition. Transitions with Ising symmetry classes, however, are not affected by this reasoning. Other studies\cite{Belitz1,Kirkpatrick,Belitz2,Karahasanovic,Conduit,Pedder,Kruger} found similar results indicating that fluctuation effects lead to competing instabilities at the ferromagnetic QCP.

\begin{figure}[tbp]
\centerline{\includegraphics[width=\linewidth]{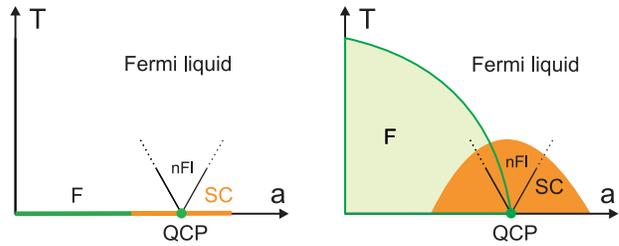}}
\caption{(Color online) Phase diagram in the space of temperature~$T$ and boson mass~$a$. The left-hand side describes the Heisenberg ferromagnet: long-range ferromagnetic order is only possible at the line $T=0$ while a $p$-wave gap appears at the QCP covering the nFl regime. The right-hand side shows the same situation for the Ising ferromagnet. Here the magnetic state has a true long-range order.}
\label{phase}
\end{figure}

In this work, we thus consider the problem from a different perspective. While it is still unclear how a complete theory that describes both the nFL physics and the transition into a competing ordered state may be constructed, we here try to understand the behavior from the point of view of the metallic site~($a\geq 0$). Using the approach of Ref.~\onlinecite{emp},  we argue that the QCP is unstable towards Cooper pairing of fermions foreshadowing the transition into the magnetic state. Due to the nature of the interaction, the gap has a triplet symmetry for ferromagnets and a singlet one for the Ising nematic transition. At the same time, true long-range order in 2D systems is of course impossible for the vector order parameters considered and correlations of such ordering in fact decay on short-range scales. Our findings are summarized in the phase diagram in Fig.~\ref{phase}.
Recent studies on the possibility of superconductivity close to the nematic QCP or to related non-Fermi-liquids using renormalization group techniques\cite{MetMross, Fitzpatrick, Torroba,Lederer,Maier} found similar results.

The paper is structured as follows: In Sec.~\ref{sec:model}, we define the model and reduce it effectively to fermions on two patches of the Fermi surface\cite{Lee,MetSachI} in order to capture the relevant low-energy physics. We derive an effective theory for the fermions alone by integrating out the bosons. We are treating the Heisenberg and Ising models
in parallel as most of calculations are the same. In Sec.~\ref{sec:mf}, we derive a set of self-consistent mean-field equations that include the feedback of the
fermionic order on the boson fluctuations. The solution of the mean-field equations reveals gaps in the Cooper channels at low temperatures with different symmetries
($p$-wave for ferromagnets and $s$-wave for the Ising nematic transition). In Sec.~\ref{sec:freeenergy}, we verify that the mean-field solutions lead to a lower free energy as compared
to unordered fermions. Fluctuations around the MF solution are studied in Sec.~\ref{sec:fluctations}. Finally, we conclude and discuss our considerations in Sec.~\ref{sec:conclusion}.

\section{Model}
\label{sec:model}


\subsection{Heisenberg ferromagnet}


We follow Metlitski and Sachdev\cite{MetSachI} and seek to describe the low energy physics near the QCP
in terms of the semiphenomenological spin-fermion model. For the Heisenberg ferromagnet,
the bosonic modes describing the fluctuations of the ferromagnetic order parameter have
the form of a real three-component vector field $\bphi = (\phi_1,\phi_2,\phi_3)$. Close to the transition,
they are governed by the Lagrangian
\begin{align}
 \mathcal{L}_{\bphi} =\frac{1}{2}  \bphi \chi_0^{-1} \bphi + \frac{g}{2} (\bphi^2)^2
 \label{defLagPhi}
\end{align}
where the bare propagator is given by the bare susceptibility
\begin{align}
\chi_0(\w,\bq) = (\w^2/c^2+\bq^2+a)^{-1}
\ .
\label{bprop}
\end{align}
In this formula, the mass term  $a$ measures the distance to the QCP, $c$ denotes the spin wave velocity,
$\w=2\pi T n$ with integer~$n$ is a bosonic Matsubara frequency, and $T$ is temperature.

\begin{figure}[tbp]
\centerline{\includegraphics[width=\linewidth]{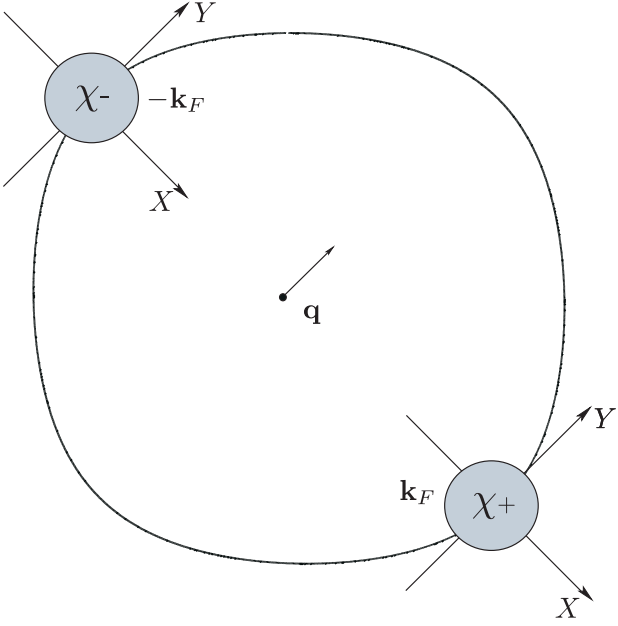}}
\caption{(Color online) Fermi surface for the ferromagnet. The colored region represents schematically the construction of the patch model. The patches belong to the Fermi momenta  $\pm \mathbf{k}_F$ that are perpendicular to the bosonic momentum $\bq$. These are the fermions interacting most strongly with such bosons.}
\label{Fig1}
\end{figure}

The Lagrangian for the spin-$\tfrac{1}{2}$ fermions~$\psi$ contains the free part,
\begin{align}
 \mathcal{L}_{\psi} = \psi^{\dagger} \left( \pd_\tau + \ep(-\ii \nabla) \right) \psi
\ ,\label{defLagFermions}
\end{align}
and the coupling to the bosons~$\bphi$,
\begin{align}
\mathcal{L}_{\psi,\bphi} = \la \psi^\dagger \bphi \bsi \psi
\ .
\end{align}
The free-fermions spectrum~$\ep(\bk)$ is assumed to lead to the Fermi surface shown in Fig.~\ref{Fig1}, and
$\la$ is the coupling constant for the interaction between fermions and bosonic ferromagnetic fluctuations.
The vector $\bsi=(\si_1,\si_2,\si_3)$ contains the three Pauli matrices for the fermion spin.
Finally, we denote by $\tau$ the imaginary time in the Euclidean field theory. The Lagrangian
\begin{align}
 \mathcal{L} &= \mathcal{L}_{\psi}+\mathcal{L}_{\bphi}+\mathcal{L}_{\psi,\bphi}
 \label{defFullLag}
\end{align}
fully determines the spin-fermion model we are now going to investigate.

The most important soft scattering processes are those in which a fermion momentum state~$\bk_F$ is scattered tangentially to the Fermi surface
to the state~$\bk_F+\bq$. The momentum~$\bq$ (or $-\bq$) of the absorbed (emitted) boson in these most important processes is thus perpendicular to~$\bk_F$.
This observation allows the construction of an effective Lagrangian\cite{Lee,Rech,MetSachI} that further reduces Eq.~(\ref{defFullLag}).
Specifically, we consider only those fermion states located in the vicinity of two antipodal patches around the opposing Fermi momenta $\bk_F$ and $-\bk_F$
of the Fermi surface, see Fig.~\ref{Fig1}, while boson momenta satisfy $\bq\perp\bk_F$.

Expanding the Lagrangian~(\ref{defFullLag}) around the momenta specified above yields
the effective patch-model Lagrangian~$\mathcal{L}=\mathcal{L}_{\chi}+\mathcal{L}_{\bphi}+\mathcal{L}_{\chi,\bphi}$
with
\begin{align}
\mathcal{L}_{\chi} &=
  \chi^{\dagger}_{+}\left(\partial_{\tau} -  \ii v_x \partial_x - v_y \partial_y^2  \right) \chi_{+}
\nonumber\\
&\qquad + \chi^{\dagger}_{-} \left(\partial_{\tau} + \ii v_x \partial_x -v_y \partial_y^2  \right) \chi_{-}
\ ,\nonumber\\
\mathcal{L}_{\bphi} &=
\frac{N}{2}\left[
  (\pd_\tau \bphi)^2/c^2+ (\pd_y \bphi)^2+ a \bphi^2
 \right]
\ , \nonumber\\
\mathcal{L}_{\chi,\bphi} &=
\la  \bphi\left( \chi^{\dagger}_{+} \bsi \chi_{+} +  \chi^{\dagger}_{-} \bsi \chi_{-}  \right)
\ , \label{Lagrangian}
\end{align}
where $v_x$ and $v_y$ are the Fermi velocity and the local Fermi surface curvature, respectively.
Also, in order to introduce a formal (though artificial) expansion parameter,
we enlarge the  number of fermion ``flavors'' by assuming $N$ copies\cite{Lee,MetSachI} of the original fermions
from Eq.~(\ref{defLagFermions}), leading to the effective fermion field~$\chi_\pm=(\psi_{1,\pm},\ldots,\psi_{N,\pm})$
The index~$\pm$ distinguishes the two patches of the Fermi surface under consideration. Note that for each flavor index~$j$, $\psi_{j,\pm}$ is still
a two-component spinor. Finally, we remark that the quartic term in Eq.~(\ref{defLagPhi}) has been neglected,
a suitable approximation for the right-hand side of the QCP, $a \geq 0$.

Before beginning our actual analysis, let us introduce a more convenient and compact notation.
For this purpose, and loosely following Ref.~\onlinecite{emp}, we extend the current $2N$-component fermion fields~$\chi_\pm$
by two more pseudo-spins, the particle-hole pseudospin (denoted by $\tau$) and
another to distinguish states in the two different patches ($\Lambda$). Working
with particle and hole-type states on a same footing is especially convenient in situations
where superconducting pairing emerges, as will be the case in the present study.
We thus define the $8N$-component fermion field
\begin{align}
\Psi^t &= \frac{1}{\sqrt2}
 \Big(
 \big(
 \chi_{+}^{\ast},\ii \si_2 \chi_{+}
 \big)_\tau,
 \big(
 \chi_{-}^{\ast},\ii \si_2 \chi_{-}
 \big)_\tau
 \Big)_\Lambda
\ .
\label{spinor}
\end{align}
Here and in the following, the superscript $t$ denotes transposition of vectors or matrices.
Introducing the matrix
\begin{align}
C =  \left(
           \begin{array}{cc}
                0 & \ii \si_2 \\
                -\ii \si_2 & 0 \\
           \end{array}
    \right)_\tau\ ,
\end{align}
we define ``charge conjugation'' for both vectors,
\begin{align}
 \bar \Psi = \left(C \Psi \right)^t
 \ ,
\end{align}
and matrices,
\begin{align}
 \bar M(X,X^\prime) = C M^{t}(X^\prime,X) C^t
\ .
\end{align}
Note that this notion of charge-conjugation extends
to dependencies on imaginary time~$\tau$ and coordinates~$\mathbf{r}$, which have been combined into
$X=(\tau,\mathbf{r})$.

This allows to rewrite~$\mathcal{L}_\chi$ and $\mathcal{L}_{\chi,\bphi}$, Eq.~(\ref{Lagrangian}),
as
\begin{align}
\mathcal{L}_{\Psi} &= \bar \Psi(X) \big[
- \pd_{\tau} + \ii v_x \pd_x \Lambda_3  -v_y \pd^2_y  \tau_3
\big]
\Psi(X)
\ ,\label{defLagpsi}\\
\mathcal{L}_{\Psi,\bphi} &=
\la \bar \Psi \bsi  \Psi \bphi= - \la \  \mathrm{tr} \left[ \bsi \Psi(X) \bar \Psi(X) \right]\bphi
\ .\nonumber
\end{align}
The latter notation involving the trace over all discrete degrees of freedom of~$\Psi$
allows an easy integration over~$\bphi$.

Then, integrating the bosons out of the partition function~$Z$ leads to a theory containing
only fermions, which interact, however. We obtain
\begin{align}
Z &= \int e^{-S[\Psi]} \left\langle e^{-S_{\Psi,\bphi}} \right\rangle_{\bphi} D\Psi
\nonumber\\
  &= \int e^{-S[\Psi]-S_{\mathrm{int}}[\Psi]}  D\Psi
\label{defLagint}
\end{align}
with the interacting part described by the action
\begin{align}
S_{\mathrm{int}}[\Psi] &=-\frac{\la^2}{2}\int \sum_{ij}\chi_0^{ij}(X-X^\prime)
\nonumber \\
& \quad \times
\bar \Psi(X)\bsi_i \Psi(X)
\bar \Psi(X^\prime)\bsi_j \Psi(X^\prime) \  dX dX^\prime  .
\label{Lagint}
\end{align}
The effective space- and time-dependent interaction potential
in this formula is formed by the bare bosonic propagator
\begin{align}
\chi^{ij}_0(\w,\bq) = \frac {1}{N} \frac{\de_{ij}}{ \w^2/c^2 + q_y^2 +a}
\ .
\label{bosonpatch}
\end{align}
Power counting shows\cite{Lee, MetSachI} that at low energies the $\w^2$-term in the boson propagator becomes
irrelevant. On the other hand, the relevant effective frequency-dependency (including Landau-damping) will eventually be generated
in form of interaction-induced self-energies so that we may discard the $\w^2$-term in the beginning.

In contrast to the this term, the kinetic term~$\pd_\tau$ in the fermionic Lagrangian, Eq.~(\ref{defLagpsi}), although irrelevant by power counting as well,
should be kept\cite{Lee, MetSachI} because of the importance of the topological information about the sign of the fermion frequency and
since otherwise, clearly, the theory would lack any dynamics. Eventually, however, also this term will be dominated by an emerging
self-energy term.\cite{Lee, MetSachI, Rech}


\subsection{Ising ferromagnet and Ising nematic transition}


In case of Ising symmetry, the critical fluctuation modes near the phase transition are described by scalar boson fields~$\phi$.
In the effective patch-model, the Lagrangian for the interaction between bosons and fermions
has instead of $\mathcal{L}_{\chi,\bphi}$ in Eq.~(\ref{Lagrangian}) the form
\begin{align}
\mathcal{L}_{\chi,\phi}^{(\mathrm{IF})} &= \lambda \, \phi \left( \chi^\dagger_{+} \sigma_3 \chi_{+} +  \chi^\dagger_{-} \sigma_3 \chi_{-} \right),
\end{align}
for the Ising ferromagnet, and we use
\begin{align}
\mathcal{L}_{\chi,\phi}^{(\mathrm{IN})} = \lambda \phi \left( \chi^\dagger_{+} \chi_{+} -  \chi^\dagger_{-}  \chi_{-} \right).
\end{align}
in order to study Ising nematic transition.

In terms of the compact notation~(\ref{spinor}), we may equivalently write
\begin{align}
\mathcal{L}_{\Psi,\phi} = \lambda \,\phi \bar  \Psi \hat M \Psi.
\end{align}
where the form of the coupling matrix $\hat M$
depends on the symmetry as
\begin{align}
\hat M &=
\left\{
\begin{array}{cl}
\sigma_3 & \quad\textnormal{Ising ferromagnet}\\
\Lambda_3 \otimes \tau_3 &\quad\textnormal{Ising nematic}
\end{array}
\right.
\label{defM}
\end{align}
Integrating out the bosons as before [Eq.~(\ref{defLagint})], we obtain similarly
an effective theory of interacting fermions~$\Psi$ with the interaction
instead of Eq.~(\ref{Lagint}) given by
\begin{align}
S_{\mathrm{int}}[\Psi] &=-\frac{\la^2}{2}\int \chi_0(X-X^\prime)\ \bar \Psi(X)\hat M \Psi(X)
 \nonumber \\ &\qquad \times
\bar \Psi(X^\prime) \hat M\Psi(X^\prime)  \ dX dX^\prime\ ,
\end{align}
where $\chi_0$ is given by the diagonal components of Eq.~(\ref{bosonpatch}).

In general, our calculational approach is independent of the three symmetries considered. For this reason, we present it in detail
only for the case of a Heisenberg ferromagnet, Eq.~(\ref{Lagint}), but discuss what is specifically different in the Ising cases
afterwards.


\section{Mean-field theory}
\label{sec:mf}

Self-energy corrections and emergent (mean-field) order parameters within the fermion model~(\ref{defLagint})
represent the impact of the bosonic fluctuation modes on the fermionic dynamics. We investigate
these effects first. In a second step, we have to see about the feedback of fermion ordering and
renormalized dynamics onto the original bosons and, hence, to the effective interaction potential
of~$S_\mathrm{int}$, Eq.~(\ref{Lagint}). This self-consistent procedure is similar
to an Eliashberg approach.\cite{Rech,Chub} In Sec.~\ref{sec:fluctations}, we will study
fluctuations on top of the mean-field approximation in order to check its stability.


\subsection{Fermion mean-field theory}


We seek nontrivial mean-field solutions in the Cooper and Fock channels
and thus decouple the interaction  $S_\mathrm{int}$, Eq.~(\ref{Lagint}), for the Heisenberg Ferromagnet in the form
\begin{align}
S_{\mathrm{int}}[\Psi,Q] &\rightarrow \ii \int \mathrm{tr} \left[Q(X,X^\prime)\Psi(X^\prime)\bar \Psi(X)\right]  \ dX dX^\prime
\ .
\label{Sint}
\end{align}
The mean-field~$Q$ is generally an $8N\times 8N$ matrix in the space of discrete degrees of freedom,
which, we recall, are spin, the pseudospin distinguishing patches, particle-hole space, and additional fermion flavors.
The form of the coupling with the fermions, Eq.~(\ref{Sint}), leads to the symmetry
\begin{align}
\bar Q(X,X^\prime) = -Q(X,X^\prime) ,
\label{Qsymmetry}
\end{align}
and it is self-consistently determined by the mean-field equation
\begin{align}
Q(X,X^\prime) = -2 \ii \la^2 \sum_{ij} \chi_0^{ij}(X-X^\prime) \si_i \left\langle \Psi(X) \bar \Psi(X^\prime) \right\rangle_{\mathrm{eff}} \si_j
\ . \label{scQ}
\end{align}
Note the factor of two which arises from summing contributions of both Cooper and Fock channels.
Averaging $\langle\ldots\rangle_{\mathrm{eff}}$ is with respect to the effective action
\begin{align}
S_{\mathrm{eff}}[\Psi] = \int \bar \Psi(X) \hat H_{\mathrm{eff}} \Psi(X) \ dX
\label{Seff}
\end{align}
with
\begin{align}
\hat H_{\mathrm{eff}}  =  - \pd_{\tau} + \ii v_x \pd_x \Lambda_3  -v_y \pd^2_y  \tau_3-\ii \hat Q
\ ,
\end{align}
where the operator $\hat Q$ acts on the fields $\Psi(X)$ as
\begin{align}
\hat Q\Psi(X) = \int Q(X,X^\prime) \Psi(X^\prime) \ d X^\prime.
\end{align}
Action~(\ref{Seff}) is quadratic in fermion fields and thus naturally defines the propagator
\begin{align}
G(X,X^\prime) = -2  \left\langle  \Psi(X) \bar \Psi(X^\prime) \right\rangle_{\mathrm{eff}}
\ , \label{defG}
\end{align}
which is an $8N \times 8N$ matrix.

In the following, we assume that $Q(X,X^\prime)$ depends only on the difference of coordinates.
Then, in the Fourier representation, the self-consistency equation~(\ref{scQ}) reduces to
\begin{align}
Q(\ep,\bp) &=
\ii \la^2 T \sum_{\ep^\prime}\sum_{ij}  \int \chi_0^{ij}(\ep-\ep^\prime,\bp-\bpp)
\nonumber\\
&\qquad\qquad\times \si_i G(\ep^\prime,\bp) \si_j \frac{d \bp^\prime}{(2\pi)^2}
\ ,\label{defsc}
\end{align}
where $\ep=\pi T(2n+1)$ with integer~$n$ is a fermionic Matsubara frequency.

Expanding $Q$ into its components in spin space~$\sigma$, we obtain the decomposition
\begin{align}
Q(\ep,\bp)= \big[f(\ep,\bp)-\ep\Big] \openone + \ii S(\ep,\bp)+\ii \bD(\ep,\bp)\bsi
\label{defQ}
\end{align}
where $f$ represents a usual self-energy correction, $S$ is the $4N\times4N$ matrix
for a possible singlet gap, and $\bD=(D_1,D_2,D_3)$ a triplet gap with each component
being a $4N\times4N$ matrix in $\Lambda\otimes\tau$ and flavor spaces. As in
the usual Eliashberg treatment,\cite{Rech,Chub} we neglect corrections to the fermion
spectrum~$\ep(\bk)$ and also retain in the following only the frequency dependency
of $Q$. Symmetry~(\ref{Qsymmetry}) implies the symmetries
\begin{align}
 \bar f(\ep) &= -f(-\ep)
 \ , \nonumber\\
 \bar S(\ep) &= -S(-\ep)
 \ , \nonumber\\
 \bar \bD(\ep) &= \bD(-\ep)
 \ . \label{ressym}
\end{align}
Singlet~$S$ and triplet~$\bD$ gaps usually do not appear at the same time so that we
have to concern ourselves only with the sector that survives. Even if the present
mean-field equations predicted degenerate singlet and triplet solutions, in reality
residual interactions such as Coulomb or electron-phonon coupling, which are always
present, lift the degeneracy and generally favor the singlet gap because of
its more robust $s$-wave symmetry. For these reasons, we will treat
$S$ and $\bD$ separately.

\subsubsection{Triplet gap}

In the situation of the triplet gap, the Green's function in the mean-field equation~(\ref{defsc})
has the form
\begin{align}
G(\ep,\bp) =  \left[\ii f(\ep)\openone -v_x p_x \Lambda_3-v_y p_y^2  \tau_3 -\bD(\ep)\bsi \right]^{-1}
\ .\label{green}
\end{align}
In order to obtain in the right-hand side the pole structure in the form of  (scalar) denominators
\begin{align}
P_{\pm} = \left[f^2(\ep) + (v_x p_x \pm v_y p_y^2)^2  +d^2(\ep)\right]^{-1}
\ ,\label{Ppm}
\end{align}
we should restrict $\bD$ by the constraints
\begin{align}
\{\bD, \Lambda_3\} = \{\bD, \tau_3\} = 0
\ ,\label{relations}
\end{align}
and demand that $(\bD \sigma)^2$ be scalar. These
relations and symmetry~(\ref{ressym}) limit $\bD$ to
matrices of the form
\begin{align}
\bD(\ep) = d(\ep) \bn U
\quad
\textnormal{with}
\quad U = \mathrm{i}\Lambda_2
           \left(
           \begin{array}{cc}
                 0 & u \\
                -u^\dagger & 0
           \end{array}
         \right)_\tau.
\label{defU}
\end{align}
Herein, the real scalar function $d(\ep)$ is the amplitude of the gap,
the unit vector $\bn =(n_1,n_2,n_3)$ corresponds to the $d$~vector of a triplet superconductor,
and $u$ is a unitary $N\times N$-matrix in flavor space. ($\Lambda_2$ is the second Pauli matrix
in the space of patches.)

The gap~$\bD$ contains only off-diagonal components in particle-hole space~$\tau$.
This corresponds to the formation of triplet Cooper pairs.
Note that there is no competing charge
order as in the case of an antiferromagnetic QCP.\cite{emp}
This absence of any charge order is ultimately due to
the curvature, which has imposed the constraint $\{\bD,\tau_3\}=0$, cf. Eq.~(33).

In the physical case $N =1$, the group of $u$ reduces to $\mathrm{U}(1)$, which just corresponds to the
phase of the superconducting condensate.
Explicitly carrying out the inversion in Eq.~(\ref{green}), we find
\begin{align}
G(\ep,\bp) = -\frac 12 \left[(P_+ +P_-)\openone + (P_+ - P_-) \tau_3  \Lambda_3  \right]
\nonumber\\
\times
\left(\ii f(\ep)\openone +v_x p_x  \Lambda_3+v_y p_y^2  \tau_3 +\bD(\ep)\bsi \right)
\label{green2}
\end{align}
with $P_\pm$ defined in Eq.~(\ref{Ppm}).

Inserting Eq.~(\ref{green2}) into the self-consistency equation~(\ref{defsc}) yields
the two equations
\begin{align}
f(\ep)-\ep =
\frac{\la^2}{2} T \sum_{i}\sum_{\ep^\prime} \int  \chi_{0}^{ii}(\ep-\ep^\prime,\bp-\bpp) \nonumber\\ \times
(P_+ + P_-) f(\epp) \frac{d \bp^\prime}{(2\pi)^2}
\ ,\label{aa} \\
\bD(\ep)\bsi =
-\frac{\la^2}{2} T \sum_{ij}\sum_{\ep^\prime} \int \chi_{0}^{ij}(\ep-\ep^\prime,\bp-\bpp) \nonumber\\ \times
(P_+ + P_-)   \si_i \bD(\ep^\prime) \bsi \si_j \frac{d \bp^\prime}{(2\pi)^2}
\ .\label{aD}
\end{align}
We are tackling their solution in Sec.~\ref{ssec:solution}.

\subsubsection{Singlet gap}

The singlet gap~$S$ behaves differently from the triplet gap~$\bD$
under charge conjugation, cf. Eqs.~(\ref{relations}). This different
symmetry behavior constraints the matrix~$S$ to
\begin{align}
S(\ep) = b(\ep) W
\quad \textnormal{with}
\quad
W = \Lambda_1\left(
            \begin{array}{cc}
                0 & w \\
                 w^\dagger & 0
            \end{array}
          \right)_\tau
\ ,
\label{defW}
\end{align}
where $b(\ep)$ is the (real) amplitude and $w$ a unitary $N\times N$ matrix in flavor space.
In particular, we find that $S$ corresponds to a $s$-wave superconducting
pairing gap.

Instead of Eq.~(\ref{aD}), the second equation in the set of self-consistency equations
in the singlet case is
\begin{align}
S(\ep) &=
-\frac{\la^2}{2} T \sum_{i}\sum_{\ep^\prime} \int \chi_{0}^{ii}(\ep-\ep^\prime,\bp-\bpp)
\nonumber\\
&\qquad\qquad \times (P_+ + P_-)\ S(\ep^\prime) \frac{d \bp^\prime}{(2\pi)^2}
\ .\label{aS}
\end{align}
Because the singlet spin space is trivial, the Pauli matrices have dropped out. A quick inspection reveals that
for the Heisenberg ferromagnet, the singlet solution vanishes as the signs in the two sides of Eq.~(\ref{aS}) are different. In the triplet case,
the anticommutation relations of the there present Pauli matrices heal this ``fault.''
For the Ising nematic transition, however, it will be the singlet component that prevails, see below.


\subsection{Feedback on bosons}


The relevant dynamics of the bosonic fluctuations is generated only by their coupling to the fermions.\cite{Lee,MetSachI}
If the fermions have developed a pairing order, this order should then also have an impact on the
bosons. Thus, for a complete self-consistency, we have to consider the feedback of the
(ordered) fermions on the boson propagator, which itself determines the effective fermion-fermion
interaction.

\begin{figure}[htb]
\centerline{\includegraphics[width=\linewidth]{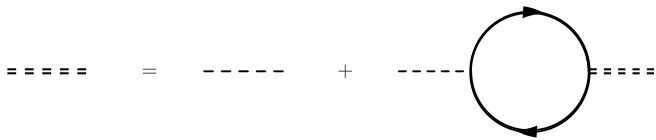}}
\caption{Effective bosonic propagator (double-dashed lines)
as a result from coupling the bare bosons (dashed lines) to the
possibly ordered fermions (solid lines).}
\label{Figboson}
\end{figure}

In practice, we obtain the self-consistently renormalized bosonic propagator~$\chi_{\mathrm{eff}}^{ij}$ by
summing the usual series in bubble diagrams, see Fig.~\ref{Figboson}. Analytically, the result is given by
\begin{align}
[\chi_{\mathrm{eff}}^{ij}(\w,\bq)]^{-1} = [\chi_0^{ij}(\w,\bq)]^{-1}-\Pi^{ij}(\w,\bq),
\end{align}
with the boson self-energy
\begin{align}
\Pi^{ij}(\w,\bq)  &= -\frac{\la^2}{2} T \sum_{\ep}
 \label{defPi}\\
 &\times \int \mathrm{tr} \left[ G(\ep,\bp)\si_i G(\ep+\w,\bp+\bq)\si_j\right] \frac{d \bp}{(2\pi)^2}
\ .\nonumber
\end{align}
The fermionic Green's functions in this equation include the mean-field $Q$.
On the other hand, the effective boson propagator $\chi_{\mathrm{eff}}^{ij}$ including
the self-energy $\Pi$ should replace $\chi^{ij}_0$ in Eqs.~(\ref{aa}), (\ref{aD}), and (\ref{aS})
so that we arrive at a completely self-consistent system of mean-field equations.

In a general situation of a nonzero triplet gap~$\bD$, Eq.~(\ref{defU}), off-diagonal elements in $\chi_{\mathrm{eff}}^{ij}$
make calculations rather cumbersome. If we assume an ordering vector~$\bn=(1,0,0)$, however,
there are nonzero components only along the diagonal and we find
\begin{align}
\Pi_{\w,\bq}^{ij}  = -\gamma N \frac{\delta_{ij}\Omega^{ii}(\w)}{|q_y|}
\end{align}
with
\begin{align}
\gamma = \frac{\la^2}{4\pi v_x v_y}\ .
\end{align}
The functions $\Omega^{ii}(\w)$
are given by $\Omega^{11}=\Omega^{-}$ and $\Omega^{22}=\Omega^{33}=\Omega^{+}$ where
\begin{align}
\Omega^{\pm}(\w) &= \pi T \sum_{\ep}
\nonumber\\
 \times\Big( 1-&\frac{  f(\ep)f(\ep+\w)\pm d(\ep) d(\ep+\w)}{\sqrt{f^2(\ep)+d^2(\ep)}\sqrt{f^2(\ep+\w)+d^2(\ep+\w)}} \Big)
\ .
\label{Ompm}
\end{align}
In the absence of the gap, they simply reduce to~$\Omega^{ii}(\w)=|\w|$.


\subsection{Solution of the mean-field equations}
\label{ssec:solution}


Let us now try and solve the mean-field equations~(\ref{aa}) and (\ref{aD}) including the renormalization
of the boson propagator for the triplet gap~$\bD$, Eq.~(\ref{defU}), in the case of the isotropic ferromagnet.
Clearly, the unitary matrix~$u$ drops out of the equation so that there are degenerate solutions
for all~$u$ in the unitary group~$\mathrm{U}(N)$. Rotational symmetry~[$\mathrm{O}(3)$] implies
degeneracy also for the normal vector~$\mathbf{n}$, which we thus choose to be $\bn = (1,0,0)$
for convenience. The full saddle-point manifold of degenerate extrema is thus $\mathrm{U}(N)\times S^2$,
where $S^2$ is the $2$-sphere.

Defining the intrinsic energy scale of the system
\begin{align}
\Gamma = \left(\frac{\la^2}{3\sqrt{3}N\ga^{1/3}v_x}\right)^3
\ ,\label{Gamma}
\end{align}
we measure all quantities of dimension energy in terms of $\Gamma$, i.e.,
$\bar T = T/\Gamma$, $\bar \ep = \ep/\Gamma$, $\bar f= f/\Gamma$, $\bar d= d/\Gamma$, and $\bar \Omega = \Omega/\Gamma$.
Then as we integrate over momenta in Eqs.~(\ref{aa}) and~(\ref{aD}), these equations together with Eq.~(\ref{Ompm}) lead at criticality ($a=0$),
to a set of fully universal equations:
\begin{align}
\bar f(\bar \ep)-\bar \ep  &= \bar T \sum_{\bar \ep^\prime,i}
\frac{\left|\bar \Omega^{ii}(\bar \ep-\bar \ep^\prime)\right|^{-1/3}\bar f(\bar \ep^\prime)}{\sqrt{\bar f^2(\bar \ep^\prime)+\bar d^2(\bar \ep^\prime)}}\ ,
\label{anr}\\
\bar d(\bar \ep)\si_1 &= -\bar T \sum_{\bar \ep^\prime,i} \si_i \si_1 \si_i \frac{\left|\bar \Omega^{ii}(\bar \ep-\bar \ep^\prime)\right|^{-1/3}\bar d(\bar \ep^\prime)}{\sqrt{\bar f^2(\bar \ep^\prime)+\bar d^2(\bar \ep^\prime)}}\ ,
\label{Dnr}
\\
\bar \Omega^{ii}(\bar \w) &= \pi \bar T \sum_{\bar \ep}
\label{anW}\\
 \times\Big(1-&\frac{\bar f(\bar \ep)\bar f(\bar \ep+ \bar \w) \pm \bar d(\bar \ep)\bar d(\bar \ep+ \bar \w)}{\sqrt{\bar f^2(\bar \ep)+\bar d^2(\bar \ep)}\sqrt{\bar f^2(\bar \ep+\bar \w)+\bar d^2(\bar \ep+\bar \w)}}   \Big)
\ .\nonumber
\end{align}
We remark that at finite temperatures~$T$, the sum over frequencies contains a divergent term for $\ep = \epp$.
This divergency only
appears as a result of neglecting the self-interactions of the bosons, the $(\bphi^2)^2$-term in Eq.~(\ref{defLagPhi}),
which at finite temperatures leads to finite mass $a(T)$ in the boson propagator, thus making the problematic term regular. At $T=0$,
this problem does not occur at all since frequency integrals converge.
We note that the issue of a divergent $\ep = \ep^\prime$ contribution does not arise\cite{Chub} in
the equations for the actual physical gap.\cite{footnote1}

\begin{figure}[tbp]
\centerline{\includegraphics[width=\linewidth]{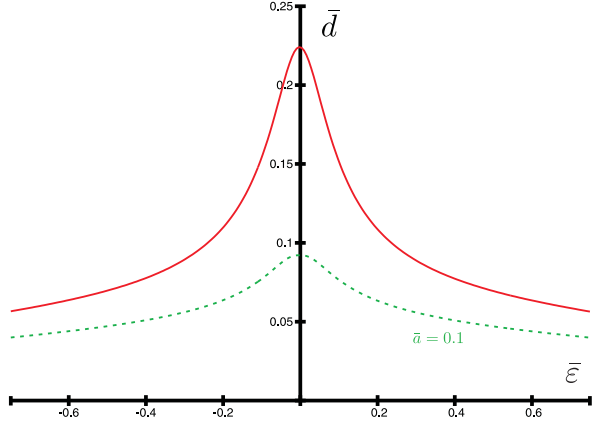}}
\caption{(Color online) Gap~$\bar{d}$ as a function of a dimensionless frequency~$\bar{\ep}$
at criticality ($a=0$) and a little to the right of the QCP ($\bar{a}=0.1$). Temperature
has been chosen to be $\bar T = 0.001$.
The gap has a maximum close to zero frequency. All energies are measured in units of~$\Gamma$, Eq.~(\ref{Gamma}).}
\label{Figgap}
\end{figure}

\begin{figure}[tbp]
\centerline{\includegraphics[width=\linewidth]{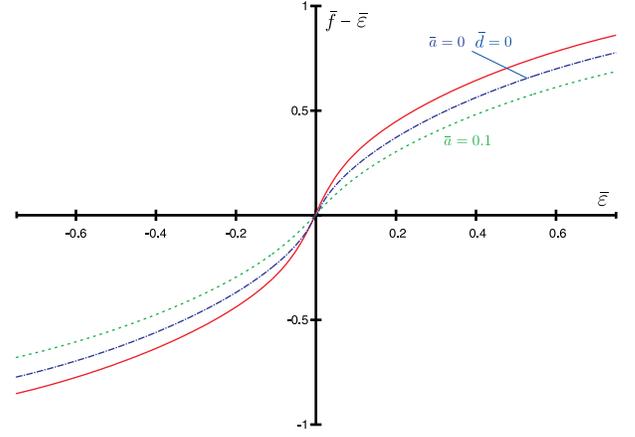}}
\caption{(Color online) Self-energy $\bar{f}-\bar \ep$ in units of~$\Gamma$, Eq.~(\ref{Gamma}), as a function of a dimensionless frequency~$\bar{\ep}$ at temperature $\bar T = 0.001$
at the QCP ($a=0$) and a little away from it ($\bar{a}=0.1$) in the presence of a nonzero gap~$d$ as
well as, for comparison, the pure self-energy solution without gap, which energetically is less favorable.}
\label{Figgap2}
\end{figure}

Equations~(\ref{anr})--(\ref{anW}) permit a solution with zero gap, $d=0$, which corresponds to the well-known\cite{Lee,MetSachI,Rech,Chub,Altshuler}
self-energy correction with $\bar f -\bar \ep \propto |\bar\ep|^{2/3}$ at low temperatures and $\bar f\propto \bar\ep$ for large $T$.
However, at low temperatures $T<T_c$ with $T_c \approx 0.07\ \Gamma$, numerical simulations show that
solutions with nonzero gaps $\bar d$ become possible and, as we show in Sec.~\ref{sec:freeenergy}, will turn out the energetically preferred ones.
At the same time, the usual self-energy terms, represented by~$f$, deviate from the gapless situation at small frequencies.
Typical frequency dependencies of $f$ and $d$ are plotted in Figs.~\ref{Figgap} and~\ref{Figgap2}.


\begin{figure}[tbp]
\centerline{\includegraphics[width=\linewidth]{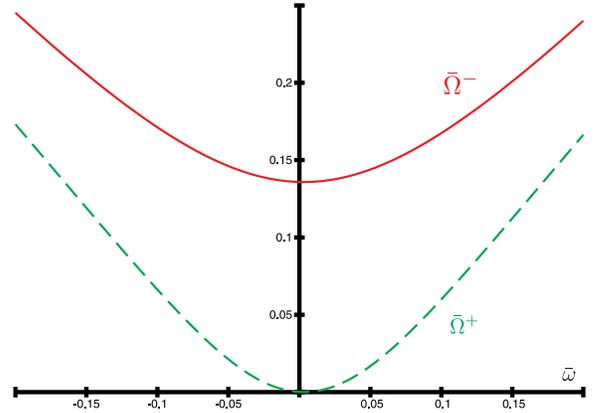}}
\caption{(Color online) Boson self-energies $\bar \Omega^{\pm}$ as a function of the frequency $\bar \omega$ at temperature $\bar T = 0.001$ and at
the QCP. At small frequencies, the gap~$d$ precludes the usual Landau damping from setting in. The component $\bar \Omega^{-}$ even acquires itself
a gap $\sim d^2$.}
\label{Figgapboson}
\end{figure}

Figure~\ref{Figgapboson} shows that also the boson self-energy shows a different than usual behavior in the gapped phases.
Moreover, its ``moment'' $\bar \Omega^{11}=\bar \Omega^{-}$, cf. Eq.~(\ref{Ompm}), in the direction
of the normal vector~$\bn=(1,0,0)$ becomes gapped at zero frequency as $\bar\Omega^{-}(\bar{\w})$ scales with $d^2$ for small $|\bar\w|$.
In contrast, the other two moments $\bar \Omega^{22}=\bar \Omega^{33}=\bar \Omega^{+}$ do not have such a gap.
For large~$\w$, these functions approach the linear regime as in the case of $d=0$.

A little away from the QCP, where $a>0$ but small, Eqs.~(\ref{anr}) and~(\ref{Dnr})
have to be modified as the last integration over momentum can no longer be performed easily.
Numerically, though, a similar study as before is possible as we replace
\begin{align}
|\bar \Omega^{ii}(\bar\ep -\bar\ep^\prime) |^{-\frac{1}{3}}
&\rightarrow
|\bar \Omega^{ii}(\bar\ep -\bar\ep^\prime) |^{-\frac{1}{3}} J\big(\bar a [\bar \Omega^{ii}(\bar\ep -\bar\ep^\prime)]^{-2/3}\big)
\end{align}
where
\begin{align}
J(\alpha) =\frac{3\sqrt{3}}{2\pi} \int_0^{\infty} \ \frac{x}{1+x^3+x \alpha}  \ d x
\end{align}
and $\bar a = (\gamma\Gamma)^{-2/3}a$ is the dimensionless bosonic mass.

For finite values $a>0$, the gap~$d$ is increasingly lowered (at $T$ fixed), see Fig.~\ref{Figgap}, and will eventually
vanish at a critical value~$a_c$.\bigskip


\subsection{Ising ferromagnet and Ising nematic transitions}


In order to analyze the situation around the Ising ferromagnetic and Ising nematic transitions,
the mean-field equations derived in the preceding section for the isotropic ferromagnet have to be adapted.
Specifically, instead of Eq.~(\ref{defsc}), the starting point should be
\begin{align}
Q(\ep,\bp) &=
\ii \la^2 T \sum_{\ep^\prime} \int \chi_{\mathrm{eff}}(\ep-\ep^\prime,\bp-\bpp)
 \nonumber\\
 &\qquad\qquad \times
\hat M G(\ep^\prime,\bp) \hat M \frac{d \bp^\prime}{(2\pi)^2}
\ ,\label{Isingmf}
\end{align}
where the matrix~$\hat{M}$, Eq.~(\ref{defM}), has to be chosen according to whether we
investigate the Ising or Ising nematic transition.

For the Ising transition, $\hat{M}=\sigma_3$ and by the same argument as in the Heisenberg case, there is no singlet solution, cf. the discussion of Eq.~(\ref{aS}).
For a triplet solution, cf. Eq.~(\ref{aD}), we need to require that $\si_3\bD\bsi\si_3=-\bD\bsi$ for a non-vanishing solution. (Otherwise, the two sides of
the mean-field equation would have different signs.) Thus, in the Ising case, we conclude that the $D_3$ component of~$\bD$ has to vanish identically.
As a result for the triplet order parameter, we obtain
\begin{align}
\bD &= b(\ep)\bn^{(\mathrm{IF})}U
\end{align}
with an effectively two-component unit vector~$\bn^{(\mathrm{IF})}=(n_1,n_2,0)$. This implies a smaller saddle-point manifold of $\mathrm{U}(N)\times S^1$.
For convenience, when writing explicit mean-field equations, we choose $\bn^{(\mathrm{IF})} = (1,0,0)$.

For the Ising nematic transition, the coupling matrix is $\hat M = \Lambda_3 \otimes \tau_3$
and does not affect the physical spin~$\sigma$. Since $\hat M U \hat M = -U$ and $\hat M W \hat M =-W$, cf. Eqs.~(\ref{defU}) and~(\ref{defW}),
both singlet and triplet pairing instabilities should be formally allowed solutions to the mean-field equations. However, as we argued after Eq.~(\ref{ressym}),
the singlet pairing should be the more robust one and prevail. Developing the mean-field theory for
the singlet pairing order parameter
\begin{align}
S(\ep) = b(\ep)W
\end{align}
based on Eq.~(\ref{aS}) is completely analogous to the scheme in the preceding section.

The mean-field equations for the pairing instabilities around both the Ising and the Ising nematic transitions
are given by
\begin{align}
\bar f(\bar \ep)-\bar \ep &=
\bar T \sum_{\bar \ep^\prime} \frac{ |\bar\Omega^+(\bar \ep - \bar \ep^\prime)|^{-1/3} \bar f(\bar\ep^\prime)}{\sqrt{\bar f^2(\bar \ep^\prime)+\bar b^2(\bar \ep^\prime)}}
\ ,\label{aising}\\
\bar b(\bar \ep)  &= \bar T \sum_{\bar \ep^\prime} \frac{|\bar\Omega^+(\bar \ep - \bar \ep^\prime)|^{-1/3}\bar b(\bar\ep^\prime)}{\sqrt{\bar f^2(\bar
\ep^\prime)+\bar b^2(\bar \ep^\prime)}}\ .
\end{align}
Note that only the (gapless) component $\bar\Omega^+$, cf. Eq.~(\ref{Ompm}), contributes to the effective boson propagator.
We emphasize that even though the gap equation are the same, the symmetry of the gap is different for both cases. For the Ising case,
$b$ is the amplitude of a two-component triplet gap, whereas this is a singlet gap in the nematic case. Numerical simulation
of the mean-field equations predicts a critical temperature of $T^{(\mathrm{IF/IN})}_c \approx 0.12\ \Gamma$, below
which superconducting triplets or singlets appear at the Ising or Ising nematic transition, respectively.



\subsection{Free energy}
\label{sec:freeenergy}


In order to complete the mean-field study, we should check that the non-trivial solution involving an emergent
pairing gap in either the triplet (Heisenberg, Ising) or singlet (Ising nematic) channel is energetically
preferable to the unordered (non-Fermi-liquid) state. Since a direct calculation
of the free energy~$F=-T\ln Z$ is difficult, we begin with the derivative with respect
to the coupling constant~$\lambda^2$,
\begin{align}
\frac{d \ln{Z[\Psi]}}{d \la^2} = - \frac{1}{\la^2} \frac{\int S_{\mathrm{int}}[\Psi] \exp\left\{-S[\Psi]  \right\} D\Psi}{\int \exp\left\{-S[\Psi]  \right\} D\Psi}
\ ,
\end{align}
and eventually recover the free energy as
\begin{align}
F = -T \int_0^{\lambda^2}\frac{d \ln{Z[\Psi]}}{d \la^2} d\la^2 + F_0
\ .
\end{align}
Here, $F_0$ is the free energy of the noninteracting system.

In zeroth other in~$\lambda^2$,
\begin{align}
\frac{d \ln{Z}}{d \la^2} &= -\frac 14 T V  \sum_{\ep,\w} \int \chi_{0}^{ij}(\w,\bq)
\label{Free}\\
&\quad \times \mathrm{tr} \left[ G_0(\ep,\bp) \si_i G_0(\ep+\w,\bp+\bq) \si_j \right] \frac{d \bp \, d\bq}{(2\pi)^4}
\ , \nonumber
\end{align}
where $V$ denotes the volume of the system.
In order to obtain the full solution, we have to sum the relevant series of diagrams. Replacing
the bare propagators for fermions and bosons by the ones we self-consistently derived
in the preceding sections, however, automatically carries out the relevant summations.
Thus, we evaluate Eq.~(\ref{Free}) with the effective propagators~$G$
and~$\chi_{\mathrm{eff}}$ for both the solution with finite gap and the zero-gap solution.
If the difference $\Delta F= F_{d\neq 0}-F_{d=0}$ becomes negative, the system will undergo the
transition into the pairing-gapped phase.

Evaluating the momentum integrals, we find
\begin{align}
 \Delta F &= \Bigg[V\int_0^{\lambda^2}\frac{\ga^{1/3}\Gamma^{4/3}S_x}{3\sqrt{3}\la^2} d\lambda^2\Bigg]
 \nonumber\\
 &\qquad\times \bar{T}\sum_{\bar{\w},i}\left([\bar{\Omega}^{ii}(\bar{\w})]^{1/3}-|\bar{\w}|^{1/3}   \right).
 \label{free2}
\end{align}
with $S_x= \int(d q_x/2 \pi)$ given by an ultraviolet momentum cutoff.
Measuring $\Delta F$ in units of the positive (nonuniversal) prefactor in square brackets in Eq.~(\ref{free2}), we find
a universal result. In particular, the sign of the free energy difference $\Delta F$ is determined by the universal part in Eq.~(\ref{free2}).
%

\begin{figure}[tbp]
\centerline{\includegraphics[width=\linewidth]{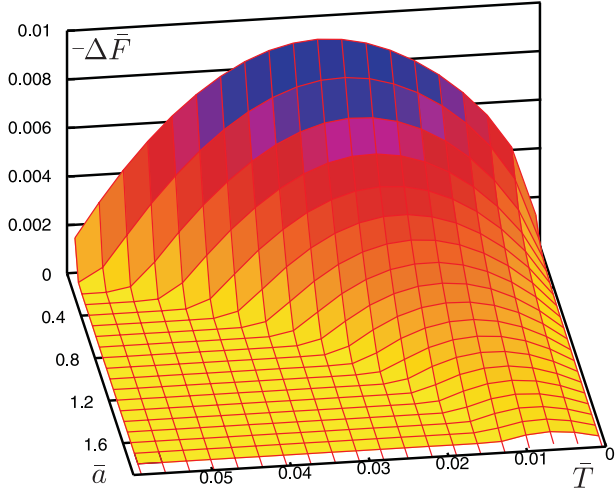}}
\caption{(Color online) Free energy difference $-\Delta F$ measured in units of
the non-universal prefactor in Eq.~(\ref{free2}) as a function of temperature $\bar T$ and distance~$\bar a$ to the QCP, both measured in units of~$\Gamma$, Eq.~(Gamma). }
\label{Figfree}
\end{figure}

The numerical evaluation of $\Delta F$ based on the solutions to the mean-field equations shows that the opening
of the gap is always energetically favorable, $\Delta F<0$. A plot of $\Delta F$ as a function
of temperature and boson mass~$a$ is shown in Fig.~\ref{Figfree}.

Without any difficulty, this result is extended
to the situation of Ising and Ising nematic transitions, where close to the QCP
superconducting pairing becomes the preferred state.

\subsubsection*{Free energy for slowly fluctuating fields}

In order to study the effects of fluctuations of the order parameter~$\bD$ (for the
isotropic case) along or perpendicular to its saddle-point manifold,
we need to generalize the free energy~(\ref{free2}) to
include (slow) fluctuations in space and imaginary time. The sought-after free energy functional should satisfy
the relation
\begin{align}
\frac{d F}{d \la^2} &= \frac {T}{4} \int\chi_{\mathrm{eff}}^{ij}(X-X^\prime)
\nonumber\\
&\qquad \times \mathrm{tr} \left[ G(X,X^\prime)\si_iG(X^\prime,X)  \si_j \right] dXdX^\prime
\ ,\label{diffF}
\end{align}
cf. Eq.~(\ref{Free}).

In analogy with Eq.~(\ref{defQ}), we split
the mean-field~$Q$ into self-energy terms and the order parameter but otherwise
keep~$Q(X,X')$ general,
\begin{align}
Q(X,X') = a(X,X') \openone + \mathrm{i}\bD(X,X')\bsi
\ .
\end{align}
Here, $a$ is a scalar and $\bD$ the order parameter including
fluctuations around the mean-field. Analogously, we split
the Green's function~$G(X,X')$, Eq.~(\ref{defG}),
\begin{align}
G(X,X^{\prime}) = A(X,X^{\prime}) \openone + \bB(X,X^{\prime}) \bsi
\ .
\end{align}
The mean-field equations locally relate the quantities introduced above to
each other,
\begin{align}
a(X,X^{\prime}) &=  \ii \la^2  \chi_a(X-X^\prime) A(X,X^{\prime}),
\label{mfa}\\
\bD (X,X^{\prime}) &= \la^2 \chi_{\bD}(X-X^{\prime}) \bB(X,X^{\prime})
\ .
\label{mfaD}
\end{align}
Here, interested in slow fluctuations, we do not assume any particular choice for $\bn$.
Thus, the structure of the equations has to take
into account nondiagonal bosonic self-energies~$\Pi^{ij}$, cf. Eq.~(\ref{defPi}). This leads to the
effective interaction couplings in Eqs.~(\ref{mfa}) and~(\ref{mfaD}) that are defined
as
\begin{align}
 \chi_a &= \sum_i \chi^{ii}_{\mathrm{eff}}
 \ ,\nonumber\\
 \chi_{\bD}^{ij} &=\chi_a\delta_{ij}-2\chi_{\mathrm{eff}}^{ij}
 \ .
\end{align}
With all these definitions at hand, we write the functional
\begin{align}
\frac{\tilde F}{T} &= \frac{1}{4} \int \mathrm{tr} \big[-\ii A(X,X^\prime) a(X^\prime,X)
\nonumber\\
&\qquad\qquad+ \bB(X,X^\prime)\bD(X^\prime,X)   \big]dX dX^\prime
\label{freefunctional}\\
&\quad -\frac{1}{2}\int \mathrm{tr \,ln} \left[\hat H_0 +  \ii a(X,X^\prime)- \bD(X,X^\prime)\bsi   \right]dXdX^\prime
\ ,\nonumber
\end{align}
where $\hat H_{0}  =  - \pd_{\tau} + \ii v_x \pd_x \Lambda_3  -v_y \pd^2_y  \tau_3$.
Differentiating $\tilde{F}$ with respect to $\lambda^2$, we almost arrive
at Eq.~(\ref{diffF}). The only discrepancy is that instead of the boson propagator~$\chi_{\mathrm{eff}}^{ij}$,
the derivative $d\tilde{F}/d\lambda^2$ holds the expression
\begin{align}
\tilde \chi_{\mathrm{eff}}^{ij} = \frac{d ( \la^2\chi_{\mathrm{eff}}^{ij})}{d \la^2}
\ .
\end{align}
However, since $\tilde \chi_{\mathrm{eff}}^{ij}$ and $\chi_{\mathrm{eff}}^{ij}$
share the same asymptotic behavior both in the ultraviolet and the infrared limits, we
believe that the approximation $\tilde \chi_{\mathrm{eff}}^{ij}\simeq\chi_{\mathrm{eff}}^{ij}$,
although not globally controlled by a small parameter, should not qualitatively alter
the physics. For this reason and simplicity, we adopt $\tilde{F}$, Eq.~(\ref{freefunctional}) as qualitatively correct
functional for the free energy.

Physically most relevant are the fluctuations of the order parameter~$\bD$.
The starting point for their study in the next section will thus be the
reduced free energy
\begin{align}
\frac{F}{T} = \frac{1}{4} \int \mathrm{tr} \left[\bB(X,X^\prime)\bD(X^\prime,X)   \right]dX dX^\prime
\nonumber\\
-\frac{1}{2}\int \mathrm{tr \,ln} \left[\hat H_0-\bD(X,X^\prime)\bsi \right]dXdX^\prime
\ .\label{finalfree}
\end{align}
We remark that in a completely analogous way, we find corresponding
free energy functionals also for the Ising and Ising nematic symmetries.


\section{Fluctuations}
\label{sec:fluctations}


In this section, we investigate the effects of fluctuations on the order parameter~$\bD$ around the mean-field solution.
Throughout most of this section, we have in mind the Heisenberg scenario while we will comment on the Ising case in Sec.~\ref{sec:rgising}.
We thus write $\bD$ as sum of the mean-field gap and its fluctuating components,
\begin{align}
\bD(X,X^\prime) = \bD_0(X-X^\prime)+\de \bD(X,X^\prime)\ .
\label{deltaD}
\end{align}
So far, we have not yet specified the nature of the fluctuations.
In fact, we can imagine different types of fluctuations according to which symmetry is broken at the mean-field level.
Below the critical temperature~$T_c$, the mean-field breaks the $\mathrm{O}(3)$ and $\mathrm{U}(N)$ rotational symmetries
associated with spin and the additional fermion flavors, respectively. Thus we expect massless excitations. Furthermore, we have to
consider fluctuations of the modulus $d$ that will lead to massive fluctuations.

Let us begin with the free energy functional~(\ref{finalfree}) and insert Eq.~(\ref{deltaD}) instead of the mean-field.
Expanding the functional to second order in fluctuations $\de \bD$ and transforming into momentum space,
we obtain a correction to the mean-field free energy of the form
\begin{widetext}
\begin{align}
\frac{\de^2F[\de \bD]}{T} &=
\frac{1}{4}T^2\sum_{\ep,\w} \int \mathrm{tr} \left[ \de\bB(\ep,\bp;\w,\bq) \de\bD(\ep,\bp;-\w,-\bq)\right]
\frac{d \bp \, d\bq}{(2\pi)^4}
\nonumber\\
&\quad +\frac 14 T^2 \sum_{\ep,\w} \int \mathrm{tr} \big[
            G\left(\ep_+,\bp_+\right) \big(\de\bD(\ep,\bp;\w,\bq)\bsi\big) \
            G\left(\ep_-,\bp_-\right) \big(\de\bD(\ep,\bp;-\w,-\bq) \bsi  \big)
            \big] \frac{d \bp \, d\bq}{(2\pi)^4}.
\label{flucs}
\end{align}
\end{widetext}
For a compact notation, we use the abbreviation $\ep_\pm = \ep \pm \w/2$ and $\bp_\pm =  \bp \pm \bq/2$. The function~$\de\bB$ is related to~$\de\bD$ by Eq.~(\ref{mfaD}).
Equation~(\ref{flucs}) is our starting point to study the effects of fluctuations.


\subsection{Massless fluctuations}


From Eq.~(\ref{defU}) we recall that the gap is described by three components, which are its modulus~$d$, a normal vector $\bn$, and the unitary matrix $u$,
cf. Eq.~(\ref{defU}).
All components fluctuate around the mean-field solution, which we choose to be at $\bn_0 = (1,0,0)^t$, $u = \openone$, and $d_\ep=d(\epsilon)$. In the harmonic
approximation, the three sectors, in particular massless and massive fluctuations, decouple, and we may study them  separately.
Let us first consider a fixed modulus $d_\ep$ and fixed ``flavor'' matrix $u$, and have a look at fluctuations in the direction $\bn$. We thus consider
\begin{align}
\de \bD(\ep;\w,\bq) = d_{\ep} U\ \de\bn_{\w,\bq},
\end{align}
as fluctuating fields with a fixed modulus ($\bn^2=1$) may be parametrized\cite{ZinnJustin} around a chosen mean-field~$\bn_0$
as
\begin{align}
\bn &= \bn_0 + \de\bn\ ,
\label{deltan}
\end{align}
where fluctuations $\de\bn$, if small, are approximately orthogonal to $\bn_0$.

Inserting Eq.~(\ref{deltan}) into the free energy~(\ref{flucs}),
we obtain
\begin{align}
\frac{\de^2F[\delta \bn]}{T} = T^2 \sum_{\w,\ep}\int \tilde{K}(\ep;\w,\bq) \de \bn_{\w,\bq} \delta \bn_{-\w,-\bq}\frac{d \bq }{(2\pi)^2}
\ ,
\end{align}
where the kernel~$\tilde{K}(\ep;\w,\bq)$ is given by
\begin{align}
 \tilde K(\ep;\w,\bq) = K(\ep;\w,\bq)- K(\ep;0,0)\ ,
\end{align}
with
\begin{align}
K(\ep,\w,\bq) =
 -N \int d^2_{\ep}\sum_{s= \pm} P_s(\ep_+,p_+)P_s(\ep_-,p_-) \nonumber\\
\times
\left[ f_{\ep_+}f_{\ep_-}
+\de_s(p_+)\de_s(p_-)+d_{\ep_+}d_{\ep_-}\right]\frac{d \bp}{(2\pi)^2}\ .
\end{align}
Here, $\de_s(p) = v_x p_x +s v_y p_y^2$ and $P_{\pm}$ has been defined in Eq.~(\ref{Ppm}).
The frequency-dependent functions $f_\ep$ and $d_\ep$ are solutions to the mean-field equations~Eqs.~(\ref{anr}) and (\ref{Dnr}).
While the single kernel $K(\ep;\w,\bq)$ separately diverges for small $|\bq|$, the effective
kernel~$\tilde K(\ep;\w,\bq)$ remains finite and, in the limit of low energies, may be expanded
for small $\w$ and $|\bq|$. This leads to the non-linear $\sigma$-model that describes the gapless fluctuations of the vector $\bn$,
\begin{align}
\frac{F[ \bn]}{T}=
  \int \big[
       a_\w \left(\pd_\tau\bn\right)^2 +
       a_x v_x^2 \left(\pd_x \bn\right)^2 +
       a_y v_y^2 \left(\pd_y \bn\right)^2 \big]\ d\tau d\br.
\label{nsim}
\end{align}
The coefficients of the three terms are
\begin{align}
\label{defawp}
a_\w &= \frac{N\rho}{v_x}\int \frac{dp_y}{2 \pi},
\\ \nonumber
  a_x &= \frac{N\eta}{v_x} \int \frac{dp_y}{2 \pi}
  \ , \quad
  a_y = \frac{N\eta}{v_x} \int \frac{dp_y}{2 \pi} p_y^2
 \ ,
\end{align}
where
\begin{align}
\label{etarho}
\eta &=
  \bar T \sum_{ \bar \ep} \frac{\bar d^2_{\bar \ep}}{(\bar f_{{\bar \ep}}^2+\bar d_{{\bar \ep}}^2)^{3/2}}
\ ,
\\ \nonumber
\rho &=   \bar T \sum_{\bar \ep} \bar d_{\bar \ep}^2  \left[  \frac{-3(\bar f_{{\bar \ep}}\bar f^{\prime}_{{\bar \ep}}+ \bar d_{{\bar \ep}} \bar d^{\prime}_{{\bar \ep}})^2}{2(\bar f_{{\bar \ep}}^2+ \bar d_{{\bar \ep}}^2)^{5/2}} \right.
\\
\nonumber
&\qquad\qquad \left. +
\frac{(\bar f_{{\bar \ep}}^{\prime 2}+\bar f_{{\bar \ep}}\bar f^{\prime\prime}_{{\bar \ep}}/2+ \bar d_{{\bar \ep}}^{\prime 2}+\bar d_{{\bar \ep}}\bar d^{\prime\prime}_{{\bar \ep}}/2)}{(\bar f_{{\bar \ep}}^2+\bar d_{{\bar \ep}}^2)^{3/2}} \right].
\end{align}
Explicit numerical evaluation of the frequency sums shows that $\eta$ and $\rho$ are both of order unity. However,
integrals over momentum~$p_y$ seem to diverge and should be cut at a proper momentum scale given by the length~$\ell$ of the Fermi surface patch.

To be specific, we see in the fermionic propagator the scaling $p_x \sim \Gamma^{1/3}\ep^{2/3}$, while the bosonic propagator
shows that $p_y \sim \gamma^{1/3} \ep^{1/3}$. Since typical energies are of order $\Gamma$, we observe
\begin{align}
 \frac{p_x}{p_y} \sim \frac{\la^2v_y}{v_x^2} \ll 1\ .
\end{align}
The last inequality determines the parameter region in which our approach is applicable.
For the length~$\ell$ of a patch, we then estimate
\begin{align}
 \ell = (\gamma \ep)^{1/3} \sim (\gamma \Gamma)^{1/3} = \frac{1}{3 \sqrt{3}}\frac{\la^2}{N v_x}.
\end{align}
Using this cutoff scale, we interpret the integrals in Eqs.~(\ref{defawp})
as
\begin{align}
 \int \frac{dp_y}{2 \pi} = \frac{\ell}{2\pi} \quad \mathrm{and} \quad  \int p_y^2 \frac{dp_y}{2 \pi} = \frac{1}{2\pi} \frac{\ell^3}{3}
 \ .
\end{align}
Finally, let us rescale the coordinates, $x=v_x \tilde x$ and $y=v_y \ell \tilde y/\sqrt{3} $, so that
the $\sigma$ model~(\ref{nsim}) becomes
\begin{align}
\frac{F[ \bn]}{T}
&= \frac{N \ell^2 v_y}{2 \pi} \int \left[ \rho\ (\pd_\tau\bn)^2 + \eta\ (\nabla \bn)^2\right] \ d\tau d \tilde \br
\ .
\label{nsimrs}
\end{align}
This $\sigma$ model is the effective low-energy theory for the gapless fluctuations of~$\bn$.

The theoretical treatment of the massless fluctuations of the unitary matrix $u$ that rotates the $N$ additional inner fermion flavors
follows the same steps and eventually leads to an effective $\si$~model with target manifold $\mathrm{U}(N)$.
We find
\begin{align}
\frac{F[ u]}{T} =  \frac{\ell^2 v_y}{2 \pi} \int \mathrm{Tr}\left[
   \rho\ \pd_\tau  u^\dagger \pd_\tau u
+  \eta\ \nabla u^\dagger \nabla u
 \right] \ d \tau d\tilde \br.
\label{nsiu}
\end{align}
Here, $\mathrm{Tr}$ is the trace over $N \times N$ matrices acting on flavors. Note that coefficients in the two $\sigma$-models~(\ref{nsimrs}) and~(\ref{nsiu})
for $\bn$ and $u$ are the same.

Concluding this section, the massless fluctuation modes around the mean-field solution are described by a combined $\mathrm{O}(3)$ and the $\mathrm{U}(N)$ nonlinear $\sigma$~model.
In Sec.~\ref{sec:rg}, we will discuss its critical behavior using the renormalization group (RG).


\subsection{Fluctuations of the modulus}


The effective free energy functional for fluctuations of the modulus can be derived
with a very similar approach. Setting $d_\ep = d_{0,\ep}+\si(\ep;\w,\bq)$, where $d_{0,\ep}$ is the mean-field value,
we investigate fluctuations of the form
\begin{align}
\de \bD(\ep;\w,\bq) =   \bn_0 U \si(\ep;\w,\bq)\ .
\end{align}
Inserting this into the free energy functional~(\ref{flucs}), we reduce
it step by step in analogy with the preceding section to the form
\begin{align}
\frac{F[\si]}{T} = \frac{N \ell^2  v_y }{2 \pi \Gamma^2} \int\left[\Gamma^2 \si^2+ \left(\frac{\pd \si}{\pd \tau}\right)^2 + (\nabla \si)^2 \right] \ d \tau d \tilde \br
\ .
\label{fluctmodulus}
\end{align}
[We here ignore coefficients $\sim 1$ of the form of those in Eq.~(\ref{etarho}).]
Fluctuations~$\sigma$ of the modulus are thus massive with mass term~$\Gamma^2$. In the prefactor, $\Gamma^2$ appears in the denominator because
$\sigma$ has the dimension energy.


\subsection{Strength of the fluctuations}


In order to investigate the limits of applicability of mean-field theory extended by the models~(\ref{nsim}), (\ref{nsiu}), and~(\ref{fluctmodulus}) for fluctuations,
let us estimate how strong fluctuations around a particular mean-field solution are.
For the gapless excitations, let us evaluate the average zero-temperature fluctuation variance. Here, the Matsubara sum is
effectively replaced by a frequency integral,
\begin{align}
 \left\langle \de\bn(0) \de\bn(0) \right\rangle &=
 \frac{4 \pi v_x }{N \ell} \int_0^\Gamma \frac{1}{\rho \w^2+\eta \left(v_x^2 q_x^2+v_y^2 \ell^2 q_y^2\right)}
 \frac{d\w \, d \bq}{(2 \pi)^3}
\nonumber\\
&\sim \frac{v_x }{N \ell} \Gamma \frac{1}{v_x v_y \ell} \sim \frac{1}{N^2}\ .
\end{align}
Thus at zero temperature, there is no generic small parameter controls the fluctuations in $\bn$ but only the (artificial) large-$N$ limit
may keep the theory under control.

At finite temperature, the contribution of the zero Matsubara frequency ($\w=0$) inevitably leads to an infrared divergency of the momentum integral,
which cannot even be healed by assuming $N\gg 1$. In fact, this just reflects the well-known result that in two dimensions,
there is no breaking of continuous symmetries at $T>0$.

Computing similarly the fluctuations of the matrix $U$, we find at $T=0$
\begin{align}
 \left\langle u(0) u^\dagger(0)  \right\rangle
\sim N \frac{v_x }{ \ell} \Gamma \frac{1}{v_x v_y \ell} \sim 1
\ .
\end{align}
Note that in contrast to fluctuations~$\de\bn$, fluctuations in~$u$ are not even suppressed in the large-$N$ limit.

Fluctuations of the modulus are massive and thus less dangerous in the infrared limit.
At zero temperature, we find
\begin{align}
\left\langle \si(0) \si(0) \right\rangle
\sim \frac{v_x \Gamma^2}{N \ell} \Gamma \frac{1}{v_x v_y \ell} \sim \frac{1}{N^2} \Gamma^2
\ .
\end{align}
Indeed, we find fluctuations in $\sigma$ suppressed but again only as $1/N^2$, yet there is no generic model parameter to control them.


\subsection{Renormalization group}
\label{sec:rg}


In order to go a little beyond the preceding section,
let us look at how the couplings of the nonlinear $\sigma$-models for the massless fluctuations in $\bn$ and
$u$ flow under the action of the renormalization group (RG). At finite temperature, only the static with respect
to imaginary time~$\tau$ component is important, yielding the effective free energy functional
\begin{align}
\frac{F}{T} = \frac{N}{t} \int \mathrm{Tr}\left[ \nabla u^\dagger \nabla u + (\nabla \bn)^2 \right] \ d\tilde{\br},
\label{freeRG}
\end{align}
with the effective temperature
\begin{align}
t = \frac{8\pi^2}{3\sqrt{3}\eta }\bar T.
\label{tbare}
\end{align}
Since the symmetries of fluctuations in~$\bn$ [$\mathrm{O}(3)$ symmetry] and the $\mathrm{U}(N)$ matrix $u$ are different, we
have to expect a different flow behavior under the RG. For this reason, we introduce two distinct effective coupling constants
for these two sectors, $1/t_u$ and $1/t_{\bn}$, which share the bare value~$1/t$, Eq.~(\ref{tbare}).

At one-loop order, we thus find the RG equations for the two nonlinear $\sigma$ models\cite{ZinnJustin}
\begin{align}
\frac{d t_u}{d\xi} = \frac{N-1}{\pi N}t_u^2, \qquad \frac{d t_{\bn}}{d\xi} = \frac{1}{2\pi N}t_{\bn}^2
\ .\label{RGUn}
\end{align}
Herein, $\xi = \log \alpha$ as we integrate out fast momenta between $\Lambda/\alpha$ and $\Lambda$.

The general behavior is independent from $N>1$. As we go to smaller momenta, the effective temperatures~$t_\bn$
and $t_u$ increase and eventually diverge in the infrared ($\xi\rightarrow\infty$), reflecting the absence of any long-range spatial order.
For $N=1$, the RG equation~(\ref{RGUn}) for fluctuations in $u$ indicates a vanishing $\beta$-function at one-loop
order. However, as it is well-known, the formation of vortices comes into play,\cite{thouless, berezinskii} and the correct RG
is given by the Berezinskii-Kosterlitz-Thouless one\cite{Jose} resulting in the appearance of quasi-long-range order.
At the same time, for angular fluctuations in $\bn$, the divergency of the RG flow of $t_\bn$ appears for all~$N$, including
the case $N=1$, showing again  the large-$N$ limit does not help to control the theory at finite temperatures.


\subsection{Ising ferromagnetic and Ising nematic transitions}
\label{sec:rgising}

In principle, we could repeat the whole procedure for the Ising symmetric models. However, results will be quite similar.
The Ising models differ from the Heisenberg one studied above as angular fluctuations of the order parameter ($\bn$) are either absent (for the Ising nematic transition)
or, in the approach above, have a vanishing $\beta$-function (for the Ising ferromagnet),  which is a general feature of the $\mathrm{O}(2)$ nonlinear $\sigma$-model.
We recall that in the physical case of $N=1$, the RG equation~(\ref{RGUn}) indicates a vanishing $\beta$-function for fluctuations in $u$ as well.

These are common features of the two-component nonlinear $\sigma$-model in two dimensions, and it is well-known that although true long-range order is impossible,
a state with quasi-long-range order emerges due to condensation of vortex and antivortex pairs. This happens at the Berezinskii-Kosterlitz-Thouless transition,\cite{thouless,berezinskii} a scenario that applies to our system in these Ising cases and leads to the quasi-long-range order at finite temperatures as indicated in
the phase diagram in Fig.~\ref{phase}.

\section{Conclusion and discussion}
\label{sec:conclusion}

We have performed a field theoretical analysis of ferromagnetic quantum-criticality for itinerant fermions in the spirit of a preceding
analysis\cite{emp} of the antiferromagnetic QCP. Using the spin-fermion model, we discussed various transition scenarios
(Heisenberg, Ising, and Ising nematic). Mean-field theory predicts a gap in the superconducting channel below
a critical temperature~$T_c$, which is either in the singlet (Ising nematic) or triplet channel (Heisenberg, Ising).
The presence of such a gap also leads to different self-energies in the boson and fermion propagators
as compared to those known from normal state studies.\cite{Altshuler,Rech,Lee,MetSachI}

Fluctuations of the gap modulus can be controlled assuming a large number~$N$ of artificial fermion ``flavors''. We conclude therefore that superconductivity with quasi-long-range order emerges for systems with Ising symmetries below a critical temperature $T_c$, see the phase diagram in Fig.~\ref{phase}. On the other hand, long-range correlations are absent in the Heisenberg ferromagnet at finite temperature for any~$N$, reflecting
the failure of $1/N$-expansions encountered in the earlier studies. However, considering the interaction between different layers in realistic systems, long-range superconducting  order might be stabilized also in the Heisenberg ferromagnet.

It is instructive to compare our results with recent theoretical works. In Ref.~\onlinecite{Lederer}, Cooper pairing close to the nematic QCP
was studied considering the interplay of nematic fluctuations and an attractive interaction in the Cooper channel. It was found that
the latter interaction determines the symmetry of the order parameter while nematic fluctuations lead to an enhancement of $T_c$. A similar
result was obtained by Ref.~\onlinecite{Maier}. For a deformed static bosonic susceptibility $\chi_0^{-1}(0,\bq) = |q_y|^{1+\epsilon}$,
Meltitski \emph{et al.}\cite{MetMross} obtained a superconducting instability covering the QCP with analytical control gained
by assuming $\epsilon \ll 1$.

On the experimental side, superconductivity close to ferromagnetism has been observed in the compounds UGe$_2$, URhGe, and UCoGe, which are Ising ferromagnets.\cite{Aoki, Aoki2}
Yet there is no unified picture for superconductivity in these compounds, which have quite different phase diagrams involving, e.g., different magnetic phases and first-order magnetic phase transition at low temperatures. Experimental evidence also links the Ising nematic transition to a superconducting instability in the iron pnictides,\cite{Chu, Fernandes} where the nematic instability is covered by a superconducting gap. Near the nematic instability, however, also a spin-density wave phase terminates, which makes the situation complicated and may even have the consequence that the superconducting gap at the QCP has a more intricate symmetry than $s$-wave determined by the residual interactions.


\section{Acknowledgments} H.M., M.E., and K.B.E. acknowledge the hospitality of the IPhT at the CEA-Saclay, where this work has begun.
K.B.E. and M.E. gratefully acknowledge financial support
from the Ministry of Education and Science of the Russian
Federation in the framework of Increase Competitiveness
Program of NUST "MISiS" (Nr. K2-2014-015). Financial
support of K.B.E. and M.E. by SFB/TR12 of DFG is gratefully
appreciated. H.M. acknowledges the Yale Prize Postdoctoral
Fellowship and C.P. acknowledges the support of PALM Labex
grant Excelcius, ARN grant UNESCOS, and COFECUB project Ph743-12.



\begin{thebibliography}{99}
\bibitem{footnote1}
The physical gap in the spectrum is given by $\Delta(\ep) = b(\ep)/Z(\ep)$ with $Z(\ep) =f(\ep)/\ep$.
\end{thebibliography}
\end{document}